\documentclass[epj, final]{svjour}

\usepackage{graphicx}
\usepackage{amsfonts} 
\usepackage{amssymb}
\usepackage{color}
\newcommand{\frep}{\ensuremath{f_{\rm rep}}}
\newcommand{\nucc}{\ensuremath{\Delta\nu_{\rm cc}}}
\newcommand{ \wmid}{\ensuremath{\omega_{\rm mid}}}
\newcommand{ \nnr}{\ensuremath{\rm nn_r}}
\newcommand{ \nnl}{\ensuremath{\rm nn_l}}

\newcommand{\kHz}{\ensuremath{\:\rm kHz}}
\newcommand{\MHz}{\ensuremath{\:\rm MHz}}
\newcommand{\GHz}{\ensuremath{\:\rm GHz}}

\newcommand{\dB}{{\ensuremath{\:\rm dB}}}

\newcommand{\nm}{\ensuremath{\:\rm nm}}

\newcommand{\cm}{\ensuremath{\:\rm cm}}

\newcommand{\Fig}[1]{Figure~\ref{#1}}

\newcommand{\fig}[1]{Fig.\,\ref{#1}}

\newcommand{\Eq}[1]{Eq.~\ref{#1}}

\newcommand{\ie}{{\it i.e.}}
\newcommand{\eg}{{\it e.g.}}

\definecolor{dkgreen}{rgb}{0,.392,0}

\begin{document}

\title{Astronomical spectrograph calibration with  broad-spectrum frequency combs}
\author{D\,.A.\,Braje\inst{1} \and M.\,S.\,Kirchner\inst{1} \and S. Osterman\inst{2} \and T.\,Fortier\inst{1} 
 \and S.\,A.\,Diddams\inst{1}}

\institute{National Institute of Standards and Technology, Time and Frequency Division, Boulder CO \and University of Colorado, Center for Astrophysics and Space Astronomy, Boulder, CO}

\date{Received: 4 March 2008}


\abstract{
Broadband femtosecond-laser frequency combs  are filtered to spectrographically resolvable frequency-mode spacing, and the limitations of using cavities for spectral filtering are considered.   Data and theory are used to show implications to spectrographic calibration of high-resolution, astronomical spectrometers.  
}

\PACS{
{42.62.Eh}{Metrological applications; optical frequency synthesizers for precision spectroscopy}\and
{95.55.-n}{Astronomical and space-research instrumentation} 
}
\maketitle

\section{Introduction} 
High-resolution astronomical spectroscopy has seen rapid advancement in the last decade. Increased light gathering and instrument stability suggest the possibility of radial velocity measurements of astronomical sources with $10$\,cm/s precision.  This precision would allow observation of terrestrial-mass planets, measurement of temporal variation of fundamental constants, and direct observation of the acceleration of the expansion of the universe.  The inherent limitations of existing astronomical wavelength standards (iodine cells and Th-Ar lamps) inhibit attaining the level of precision necessary for these measurements.
An ideal wavelength standard would provide a high-density array of uniformly spaced, constant spectral brightness emission lines, the frequency of each tied to fundamental constants or to the standard SI second.  In addition, the precision and long-term stability of this source should exceed the baseline of a high-resolution spectrograph.
Optical frequency combs based on femtosecond lasers  address all of these requirements~\cite{2007Murphy,2007Schmidt}.

The current generation of frequency combs produces an array of evenly spaced frequency modes spanning hundreds of nanometers~\cite{HallNobel,HanschNobel}.
While numerous types of femtosecond laser combs exist, the spectral range of astronomical interest from $\sim$400-2000\nm\ is well covered by two types of passively mode-locked lasers that employ mature technology~\cite{endnote1} and that have been frequency-stabilized.  The first type is based on Ti-doped sapphire, and the second class centers around Er-doped optical fiber technology.  
Ti:sapphire femtosecond lasers with  intracavity spectral broadening  centered at 800\nm\ have a typical fractional bandwidth $\Delta\lambda/\lambda\sim0.75$  (coverage from $\sim$600-1200\nm)~\cite{Ell01,2002Bartels,Fortier03}.  Still greater spectral coverage, 400-1500\nm, can be achieved by spectrally broadening the output of the Ti:sapphire laser in nonlinear microstructured optical fiber~\cite{Ranka00,Kumar02,Dudley06}.   In the near-infrared, femtosecond Er:fiber lasers are readily  broadened in highly nonlinear fiber to provide coverage from 1000-2200\nm~\cite{Nicholson03,Tauser03,Washburn04}.  
Despite sufficient spectral coverage by these lasers, the small frequency mode spacing of  $\sim$1\GHz\ for Ti:sapphire and $<$250\MHz\ for Er:fiber prohibits resolution of individual frequency modes in typical astronomical spectrographs.  

This paper discusses Fabry-Perot cavity filtering of  femtosecond-laser frequency combs, 
ultimately resulting in a wide spectrum of resolvable frequency modes with line spacing in the range of 10 to 30\GHz. 
Cavity filtering is demonstrated by use of a $\sim 1\GHz$ Ti:sapphire laser, and the effect of mirror parameters, refractive index of the cavity,  offset frequencies, mode suppression, and comb linewidth are explored.  The experimental results are compared with a theoretical model, and the implications of using a comb as a spectrographic calibration standard are considered.  


\section{Cavity Filtering Theory}
\begin{figure} \centering{
\includegraphics*[width=3.4truein]{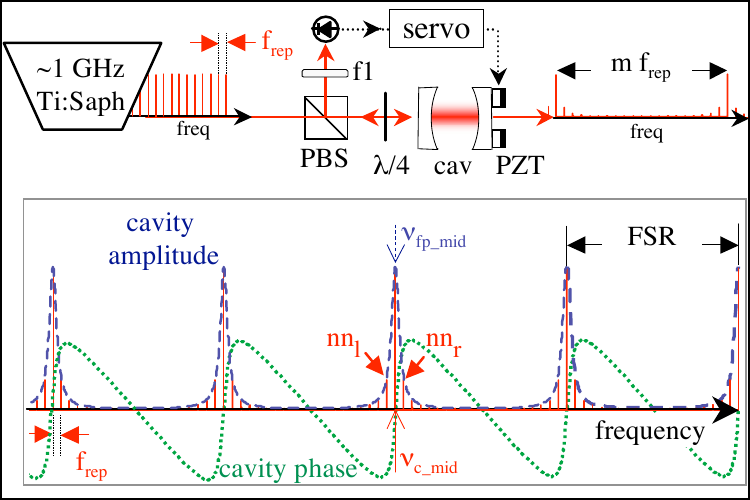}\\
}%
\caption{A \frep=\,1\GHz\ mode-locked laser is filtered to a frequency-mode spacing of $m \frep$ by use of a Fabry-Perot cavity. 
The light reflected from the cavity is picked off by a polarizing beam splitter ({\bf PBS}), passes through an optical, bandpass filter ({\bf f1}), and is the input to a PZT-based servo system controlling the length of the cavity.
 The lower section shows cavity amplitude (blue dashed) and phase (green dotted) that act on the electric field of the comb.  An offset between the cavity and the comb at the middle comb mode is defined as $\nucc=\nu_{\rm fp\_ mid}-\nu_{\rm c\_ mid}$.  The modes to the right and left of the $m^{\rm th}$ modes are defined as the right ({\bf \nnr}) and left ({\bf \nnl}) nearest-neighbor modes.}
\label{fig_defs}
\end{figure}

Using a \frep$\sim$1\GHz, Ti:sapphire comb as a prototypical laser, the basic principle of cavity filtering is illustrated in \fig{fig_defs}.  A mode-locked laser produces a comb of lines equally spaced in frequency~\cite{1999Udem}, which in current, state-of-the-art systems are spaced too tightly in frequency to be resolved by standard astronomical spectrographs.  As suggested by Sizer~\cite{1989Sizer}, a Fabry-Perot cavity may be used to spectrally select a set of modes from the comb that have a frequency spacing larger than \frep.
The length of the cavity is adjusted such that each cavity peak passes a comb mode that is separated from the previous cavity mode by  $m$ comb modes, thereby creating a comb with frequency spacing $m$ times that of the input.   The greater the mirror reflectivity, the higher the suppression of the off-resonant modes.  

This straightforward concept of using a cavity to suppress intermediate frequency modes is complicated by hundreds of nanometers of spectrum necessary for spectral calibration~\cite{1988DeVoe,2000Jones}.  While the frequency modes of the laser are equally spaced across the entire spectrum, the cavity-mode spacing is frequency dependent. Dispersive media inside the cavity as well as additional phase contributions of mirror coatings cause the frequency mode spacing or free spectral range (FSR) of the cavity to vary across the spectrum.  If the comb and cavity are well-overlapped at one spectral region, they walk off one another in other regions.  In order to maximize the bandwidth of light transmitted through a cavity while maintaining suppression of neighboring frequency modes, the mirror reflectivity $R(\omega)$, mirror phase $\phi_r(\omega)$, refractive index inside the cavity $n (\omega)$, and shift frequency $\nucc(\omega)$ between the cavity and the comb must be considered.   The offset frequency $\nucc(\omega)$ is defined as the frequency shift between the peak of cavity transmission and the nearest comb mode at a given frequency.  

\begin{figure}[b]
\centering{
\includegraphics*[width=3.4truein]{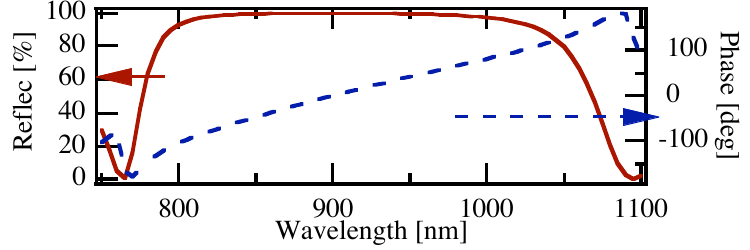}
}
\caption{Mirror coating data provided by Ramin Lalezari of Advanced Thin Films. The phase data include the linear propagation component that is subtracted for use in calculations.}
\label{fig_Mirror}
\end{figure}

In designing a large-bandwidth filter cavity that exhibits high suppression of intermediate frequencies, the most significant terms develop from the mirror coatings themselves.  For  low-group-velocity-dispersion mirrors such as those in \fig{fig_Mirror},  there is a trade-off between higher reflectivity (which increases suppression of neighboring modes) and the bandwidth over which the cavity has a uniform FSR (where the cavity can be well matched to the comb).   Even small changes in the FSR are detrimental to the usable spectral bandwidth of the cavity.   Like a Vernier scale with one metric having inconsistent spacing, the walk-off between the comb and the cavity is {\it cumulative}, with each successive $m^{th}$ comb mode slipping further from the cavity mode. 

At first inspection of the mirror reflectivity in~\fig{fig_Mirror}, it appears that the mirrors' coatings have bandwidths of a few hundred nanometers.  When accounting for mirror phase, the bandwidth over which the cavity  can filter the comb is much less than the region with acceptable reflectivity.  Assuming the mirror phase is linear in wavelength, the index inside the cavity is unity, and $R\approx 1$, the maximum mirror phase that can be tolerated while maintaining $I_{\rm out}/I_{\rm in}\geqslant 0.5$ is $|\Phi| \gtrsim\sqrt{(R-1)^2/R}$.  Note that $\Phi$ refers to the difference in mirror phase across the total usable bandwidth beyond the linear phase acquired due to normal propagation.   In actuality, the mirror phase may oscillate between positive and negative values until the cumulative phase exceeds that of the above expression.  Here we can neglect the Gouy phase shift, which has negligible  frequency dependence.  Using the above expression and the mirror phase in \fig{fig_Mirror}, a cavity bandwidth of $\sim 100\nm$ centered at 800\nm\ should be achievable.

When the cumulative mirror phase is small, such as for specially designed coatings or for metal mirrors, the index of refraction of air inside the cavity can become the dominant term affecting usable bandwidth.  Assuming a linear variation of the index as a function of wavelength within the cavity (a reasonable approximation for air in the visible spectrum) the bandwidth $\Delta \lambda$ over which comb transmission $\geqslant50\,\%$ is limited by the change in refractive index ($\Delta n$) as $\Delta\lambda\lesssim m \lambda^2\frep\sqrt{(R-1)^2/R}/(\pi c \Delta n )$, where $\lambda$ is the center wavelength of the cavity.  The filter bandwidth varies inversely with $\Delta n$; whereas, the dependence on  $\frep$ and $m$ is linear.  This somewhat counterintuitive linear relation stems from $\frep$ and $m$ being inversely proportional to cavity length.   A refractive index change of up to 1 part in $10^6$ allows an optical bandwidth of the cavity of $100\nm$.
 
The above equations give a terse understanding of the cavity limitations.  To take properly into account the cavity-comb interaction,  the system is modeled by considering the electric field of all delta-function comb lines, each spaced by a frequency \frep, interacting with an air-spaced Fabry-Perot cavity of nonabsorbing mirrors.  The steady-state, planar  or tilted-wave cavity equation is used to model the effect of the cavity on the amplitude and phase  of each comb mode~\cite{Born}:
\begin{equation}
E(\omega)=\frac{1-R(\omega)}{1-R(\omega)e^{i[2 n(\omega) \omega L/c+2 \phi_r(\omega)+\phi_{D}]}}.
\label{eq_cav}
\end{equation}
$R(\omega)$ and $\phi_r(\omega)$ account for  the coefficient of reflection and phase upon reflection of the cavity mirrors.  Because this is a planar derivation, $\phi_D$ is added empirically
 and considers all geometrical phase terms such as the Gouy phase ($\sim2\cos^{-1}[1-L/r]$ where $r$ is the radius of curvature of the cavity mirrors).  The index of refraction, $n(\omega)$,  is taken for that of air using the Ciddor equation \cite{1996Ciddor}\ for typical laboratory conditions of 24\,$^{\circ}$C, 630~Torr, a fractional humidity of 30\,\%, and 400~ppm of CO$_2$.  The cavity length $L$ is adjusted such that the filter cavity passes an integral number $m$ of comb modes through each cavity FSR.   Unless otherwise noted, all theoretical curves assume that the comb modes overlap the cavity modes; \nucc\,=\,0 at the center of the mirror coatings \wmid.

Modeling the cavity parameters by use \Eq{eq_cav} reproduces the observed spectrum in both the time and frequency domains.  Although cavity filtering is practical for increasing the temporal pulse rate~\cite{1989Sizer,1998Abedin,2004Yiannopoulos} as well as for improving performance of microwave generation by optical frequency combs~\cite{2005McFerran,2008Hollberg}, here we focus on the cavity effect of thinning  comb frequency modes.   By taking into account the amplitude and phase of the cavity from \Eq{eq_cav} on each of the incident comb frequency modes, a theoretical model of the output spectrum is possible.    

\section{Experimental setup}
\Fig{fig_defs} depicts the apparatus used in the experiments.  A broadband, Ti:Sapphire laser~\cite{2006Fortier} generates a $\sim $1\GHz\ repetition rate, octave-spanning (550\nm-1100\nm) comb.   
The frequency of the $N^{th}$ cavity mode may be written as $f_N=N \frep +f_o$.  The comb has two degrees of freedom:  the carrier-envelope offset frequency $f_o$ and the mode spacing \frep.  The repetition frequency is varied with cavity length and is easily accessible and controllable.  The offset frequency is less readily determined.  Taking advantage of the full comb spectrum which exceeds an octave of bandwidth, a group of modes at one end of the spectrum may be frequency doubled and then compared to comb modes on the other side of the spectrum.  By adjusting the pump laser power with an acousto-optic modulator, the observed beat frequency which corresponds to $f_o$ may be referenced to a known frequency.   
Both $f_o$ and \frep\ are ultimately referenced to a hydrogen maser;  the resulting spectrum consists of $\sim 10^6$ comb frequencies, equally spaced by \frep.  To more closely match the spectrum of the lasers to the reflective center of the cavity, the laser is first filtered by multiple reflections from plane mirrors of the same coating as the cavity.  


The remaining spectrum is mode-matched into a Fabry-Perot filter cavity of variable length. The cavity consists of two spherical mirrors with radii of curvature 50\cm\ and dielectric coatings of peak reflectivity 99.2\,\% centered at 910\nm.  In order to stabilize the cavity length, the light reflected from the cavity is separated by a polarizing beam splitter and is detected on a photodiode.  
A tunable, optical bandpass filter (f1 in \fig{fig_defs}) selects the portion of the optical spectrum used to lock the cavity.  The detected light creates an error signal, which maximizes the intensity of the cavity-transmitted modes in the spectral bandwidth of f1.  
This error signal drives a PZT servo system, dithered at 74\kHz, that controls the length of the cavity.   The cavity length may be scanned and then locked such that it passes $m$ comb modes per FSR.   Because the Ti:sapphire laser is not a purely TEM$_{\rm oo}$ mode, excitation of only the lowest-order cavity mode requires additional spatial filtering.  The comb teeth that traverse the cavity in the lowest-order cavity mode are detected either with a fiber-coupled fast photodetector to obtain a RF spectrum or with an optical spectrum analyzer to measure the optical spectrum.

\section{Experimental results}
\subsection{Mirror Reflectivity and Phase}
\begin{figure}[t]
\centering{
\includegraphics*[width=3.4truein]{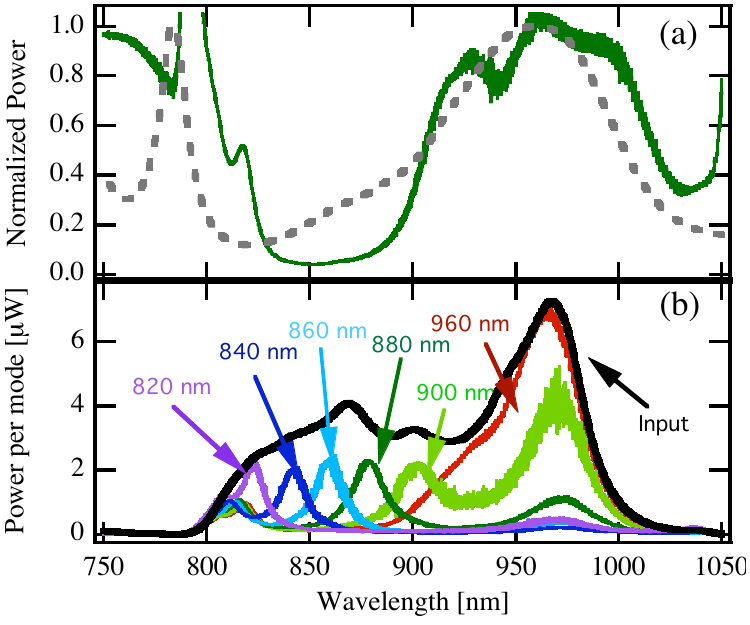}\\
}
\caption{(a) Experimental (green solid) and simulated (grey dashed) transfer function ($I_{\rm out}/I_{\rm in}$) for the $m=20$ comb modes.  Cavity length $L\approx0.75 \cm$ is locked with filter f1 centered at 980\nm. (b) Effect of cavity lock wavelength (as determined by f1) on transmission through the cavity.  Cavity-filtered comb spectra in power per mode as a function of wavelength are shown for the lock wavelength centered at 820\nm, 840\nm, 860\nm, 880\nm, 900\nm, and 960\nm.  The input spectrum has been vertically scaled by $m=20$ to account for the suppression of intermediate modes. }
\label{fig_OSASpecplot}
\end{figure}

As described above, the ideal spectrographic calibration source spans more than an octave of spectral bandwidth; however, a cavity (which is necessary to filter the dense spectral modes) has limited spectral range over which its modes overlap the evenly spaced comb modes.    Let us first consider how much spectral bandwidth can be transmitted through a cavity.  
\Fig{fig_OSASpecplot}(a) depicts the experimental (green, solid) and theoretical (grey dashed) cavity-filtered spectrum normalized to the input spectrum while the cavity is locked at 980\nm.  The experimental data are taken on a high-resolution optical spectrum analyzer with resolution bandwidth of 0.2\nm.  The cavity effectively filters nearly 100\nm\ of the input spectrum before the cavity modes walk off the filtered comb modes.  The fuzzy appearance of the transmitted spectrum results from the optical spectrum analyzer sampling the discrete transmitted comb frequencies (and the absence of neighboring modes).   The theoretical curve of $I_{\rm out}/I_{\rm in}$ of each $m^{th}$ comb mode was generated by use of experimental parameters, mirror coatings of \fig{fig_Mirror}, and \Eq{eq_cav}.  Although the mirror reflectivity bandwidth appears to cover much of the spectrum from 800-1000\nm, the usable bandwidth from the cavity is limited by the variation of the phase shift of the mirrors, as discussed above.  

The overlap of the cavity and comb modes at the locking wavelength of the cavity greatly affects the  relevant bandwidth of the cavity filter, as shown in \Fig{fig_OSASpecplot}(b).   Here the power/mode of the comb spectra is shown, and the input curve is scaled by $m=20$ to account for the filtered modes. The cavity parameters are initially optimized by use of the entire spectrum, and then different optical bandpass filters are placed in the path of the servo detector.  Optical spectra at a number of locking wavelengths are depicted.  Note that the spectral width of the transmitted modes decreases as the cavity lock point is shifted away from 960\nm.  This decrease results from the cavity-comb misalignment, or a shift in \nucc\ far away from the well aligned center modes.

\subsection{Cavity-Comb Offset} 

\begin{figure}[h]
\centering{
\includegraphics*[width=3.4truein]{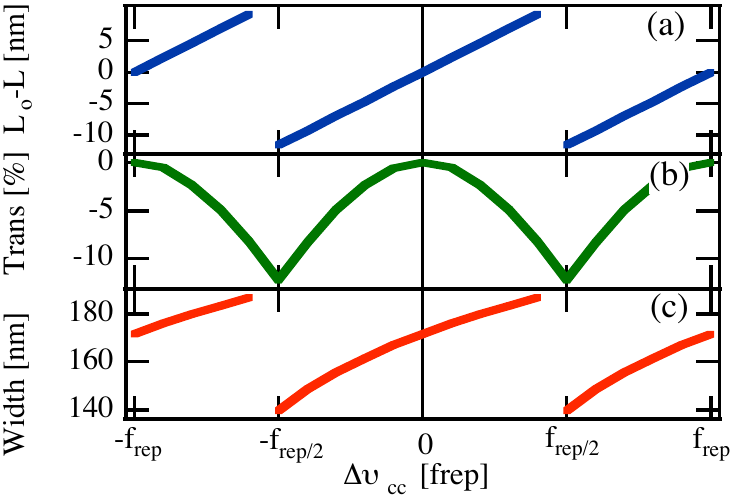}}%
\caption{Calculation of the effect of \nucc\ of the comb relative to the cavity for the mirror design of \fig{fig_Mirror}. (a)  As the offset shifts over the range of \frep (1\GHz), the cavity length  for optimum transmission and bandwidth vary linearly;  at $\nucc=\frep$, the optimum alignment point shifts to the next peak in the length scan as represented by the discontinuities. (b) Less than $15\,\%$ variation in cavity transmission in a 50\nm\ bandwidth is seen as \nucc\ is varied.  If mirrors of higher reflectivity are used, this variation increases. (c) Spectral width slightly increases as \nucc\ increases; however, note that the increase in bandwidth is at the expense of transmission.}
\label{fig_CEO}
\end{figure}

If the the offset frequency $f_o$ of the laser is adjusted (by changing the dispersion in the laser cavity) such that the Fabry-Perot cavity length is an exact integral number of comb modes, but cavity and comb modes do not overlap at \wmid\ (\ie, $\nucc=\frep/2$), it would seem that no light could be transmitted through the cavity.  Without any cavity length adjustment, the comb modes are not in line with the cavity modes, and they are largely filtered.  In practice, however,  the length of the cavity may be adjusted, very narrowly changing the cavity length away from an integral number of comb modes; in which case, the cavity length shift aligns the center frequency modes, but at the same time forces the cavity length to differ from an exact integral of comb modes.  \Fig{fig_CEO} depicts theoretical curves adjusting \nucc\ near the center of reflectivity of the mirror coatings and plots the effect on the length adjustment  of the cavity, total transmission through the center of the cavity,  and the spectral bandwidth, at $m=20$. To separate differing effects, the profiles assume a cavity in vacuum.

As \nucc\  is adjusted, the cavity length has a corresponding  linear adjustment as seen in \fig{fig_CEO}(a). This adjustment from the optimal cavity length causes a corresponding decrease in the peak amplitude transmitted through the cavity.  \Fig{fig_CEO}(b) is plotted using the experimental mirror parameters.  Higher mirror reflectivity would cause further reduction of the power through the cavity as the offset \nucc\ increases.
For maximum transmission, the offset of the laser must be tailored to the cavity.  While taking experimental data,  \nucc\ is not optimized, but the cavity length is adjusted to compensate.

It is perhaps counterintuitive that the spectral bandwidth (as defined by $I_{\rm out}/I_{\rm in}>0.5$) passing through the cavity slightly increases  as \nucc\ is increased [\fig{fig_CEO}(c)].  The broadening in the spectral width comes at the expense of peak amplitude through the cavity in the case of vacuum.  It should be noted that the spectral bandwidth may be increased further when accounting for the index of air in the cavity.  The nearly-linear increase in index somewhat balances the effect of the cavity mode's slip from the comb mode.  

\subsection{Side-Mode Suppression}
Until now, the discussion of functional cavity bandwidth was centered on cavity transmission where the $m^{th}$ comb modes have $ \textgreater 50\,\%$ transmission.  For applications such as high-precision spectrographic calibration, the definition of usable bandwidth depends more critically on the suppression of outlying frequency modes.  An ideal calibration source resolves each $m^{th}$ comb mode.  If other comb modes are not adequately suppressed, they can shift the spectral calibration causing systematic errors. 

\begin{figure}[t]
\centering{
\includegraphics*[width=3.4truein]{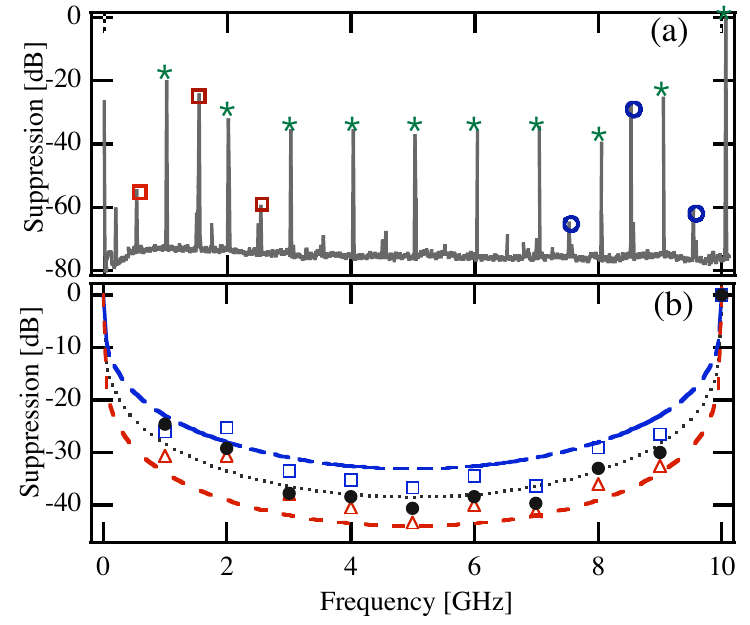}}
\caption{
(a) Heterodyne beat note of laser against filtered comb normalized to the $m^{th}$ mode.  RF spectra shows both the cumulative RF beat ($\color{dkgreen}\ast$) as well as the heterodyne ($\scriptscriptstyle{\color{red}\square}$) and conjugate heterodyne ($\color{blue} \circ$) beat notes. 
(b) Cycloid-like function of cumulative mode suppression as a function of RF frequency taken with the comb locked at different wavelengths.  Red triangles [920\nm], black circles [960\nm] and blue boxes [1000\nm] represent data taken in differing spectral regions.   The red-dashed, black-dotted, and blue-dashed lines are theoretical simulations using experimental parameters.  }
\label{fig_RF}
\end{figure}

The standard method of spectral filtering has been demonstrated by measuring the RF spectrum of comb after the filter cavity~\cite{2007Osterman}.  A fast photodiode detects the cumulative RF signal of all comb modes beating against each other.  For a \frep\,=\,1\GHz\ laser through a $m=10$ filter cavity, the data and theory are shown in \fig{fig_RF}(a) by the points marked by $\color{dkgreen}\ast$.  The first peak at 1\GHz\ results from the vector sum of each comb mode beating with its nearest neighbors; the second peak at 2\GHz\ from second nearest neighbors.  The RF signal at 10\GHz\ derives from the sum of every $10^{th}$ cavity mode.        Let us examine the 1\GHz\ beat note.  Taking into account the phase of the cavity as shown in \fig{fig_defs}, each individual 1~GHz beat note of a single comb mode against its neighbor will vary in phase from one to the next.  The cumulative RF spectrum, each beat note a vector sum, may appear to have increased side-mode suppression.
In the case where the comb and the cavity are not well aligned, the overestimate of suppression can be quite large.  \Fig{fig_RF}(b) shows the cumulative suppression measured with the cavity locked at 920\nm, where the comb and cavity are well matched, and at 960\nm\ and 1000\nm\ where the cavity-comb alignment degrades.  The theoretical curves (solid lines) take into account the vector sum of all beat notes.

A  more accurate measure of nearest neighbor suppression may be achieved by heterodyning the output of the cavity against another laser.  Rather than consider the suppression of all $m-1$ filtered comb modes, here we are concerned only with the largest side modes.
 Due to the Airy function amplitude of the filter cavity, the modes which are least suppressed are those directly neighboring the $m^{th}$ comb modes.   \Fig{fig_defs} defines the left- and right-nearest-neighbor frequency modes (\nnl\ and \nnr) at the center frequency of the spectrum.  Because \nnl\ and \nnr\ modes have the least suppression of the filtered modes (unless the cavity and comb are completely misaligned), a more meaningful definition of the usable bandwidth may be defined using the suppression of these modes. Using the amplitude from \Eq{eq_cav} for the filtered modes with respect to the $m^{th}$ modes, the maximum suppression $S$ in \dB\ of the $k^{\rm th}$ sideband at reflectivity $R$ is:
\begin{equation}
S =10 \log \left[\frac{R^2-2 \cos \left(2\pi\frac{ \nucc}{m \frep}\right) R+1}
                             {R^2-2 \cos \left[2\pi \left(\frac{\nucc}{m \frep}+\frac{k}{m}\right) \right] R+1}
              \right]
              \label{eq_suppress}
\end{equation}
where $k=\pm 1$ refers to the nearest neighbor modes.

\begin{figure}[h]
\centering{
\includegraphics*[width=3.4truein]{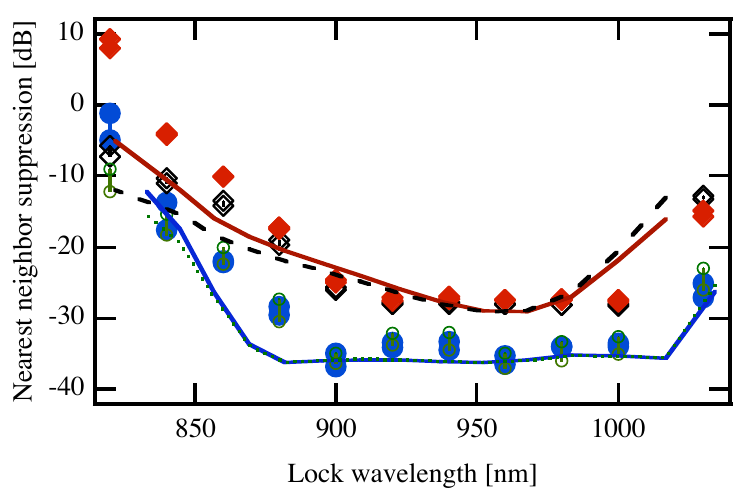}}%
\caption{ Data are plotted by measuring the suppression of the \nnl\ and \nnr\ at each of the locking bandwidths.  Open and filled diamonds are the suppression of \nnl\ and \nnr\ at $m=20$, and open and filled boxes are for $m=10$.  The solid lines represent the theoretical curves for right (solid line) and left (dashed line) nearest neighbors.}
\label{fig_heterodyne}
\end{figure}

The $\scriptscriptstyle{\color{red}\square}$ beat notes in \fig{fig_RF}(a) denote the RF frequencies resulting from a $\sim$ 4~mW cw diode laser at 960\nm\  beating against the $m^{th}$ comb tooth.   By measuring the suppression of the right and left $\scriptscriptstyle{\color{red}\square}$ beat notes against the peak, a more physical measurement of the suppression of nearest neighbors (\nnl\ and \nnr) may be obtained. 
The conjugate beat note of the laser against the $m+1$ mode is marked with  $\color{blue} \circ$.
To use the filtered comb as calibration source, it is necessary to know the suppression of intermediate modes as a function of wavelength.   
Because the heterodyne laser is not tunable across the entire bandwidth, the comb is locked by use of different frequency bands of the output cavity spectrum.    \Fig{fig_heterodyne} shows the theoretical and experimental curves  at $m=10$ (solid and filled circles) and $m=20$ (solid and filled diamonds) for the suppression of \nnl\ and \nnr.  The simulated curves take into account the full theoretical model where the length of the cavity is adjusted as a free parameter.    Both $m=10$ and $m=20$ show similar dependence upon locking wavelength; however, the maximum suppression becomes greater as the number of intermediate modes to be filtered is decreased.  
In a high-resolution spectrograph where side-mode suppression can cause an apparent shift in calibration, the limiting factor for the usable spectral bandwidth of the cavity results from both the suppression of the side modes as well as from the asymmetry of the suppressed modes.

\subsection{Filtering an Octave of Comb Bandwidth}
\begin{figure}[h]
\centering{
\includegraphics*[width=3.4truein]{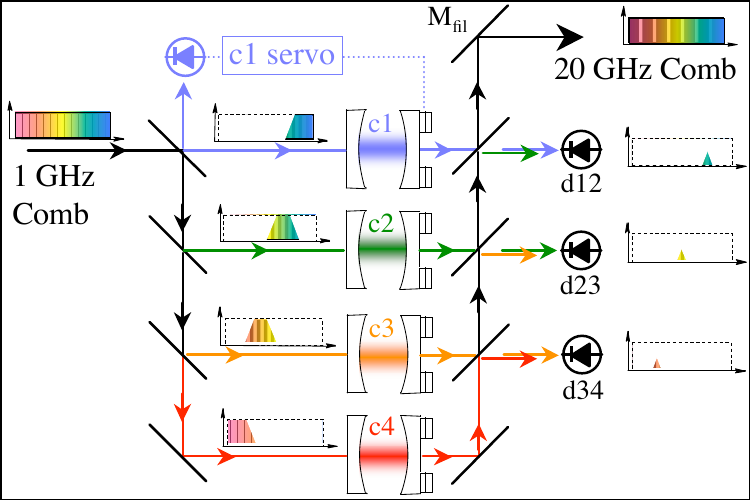}}%
\caption{Because a single cavity with mirror coatings of \fig{fig_Mirror} cannot filter a wide spectrum, the comb is spectrally divided among a number of filter cavities.  Cavities (c1-c4) process the spectrum in parallel, each cavity filtering a portion of the spectrum to the necessary frequency-mode spacing.  The length of individual cavities  is stabilized by a PZT, and the offset between the cavities (an integral of \frep) is monitored with detectors (d12, d23, d34).}
\label{fig_Multcomb}
\end{figure}

Use of a frequency comb for astronomical spectrograph calibration requires (1) a large spectral bandwidth, (2) frequency modes spaced by 10-30\GHz, and (3) high suppression ($\textgreater 27$\dB) of neighboring modes for velocity calibrations accurate to cm/s~\cite{2007Murphy}.  
These combined requirements are unphysical for a single filter cavity with our current coatings; however, multiple cavities in parallel allow coverage over the entire spectrum~\cite{2007Schmidt}. As depicted in \fig{fig_Multcomb}, the comb spectrum may be split into functional spectral bandwidths by use of dichroic mirrors.  Each cavity (c1-c4) filters the comb to $m\frep$, and the filtered combs are recombined.  
The final mirror, $\rm M_{\rm fil}$, may be designed to spectrally flatten the comb output.   In place of $\rm M_{\rm fil}$, a grating/spatial light modulator system could be used. Although the filtered frequencies of each cavity are spaced by $m\frep$, the cavities may have a frequency offset.   For example, cavity c1 may filter the $m$ modes over the blue wavelengths while cavity c2 could pass the $m+5$ modes (\ie\ both cavities have a filter number $m$, but may pick out different comb teeth). This offset, in which the $m^{th}$ teeth are filtered, can then be used to reference one cavity to the next.   In the above example, the beat note between the overlapping spectral region of c1 and c2 is $5\frep$.  

The offset of each cavity  with respect to the next  may be designed directly into the mirror coatings, or more practically, a slight change in cavity length may be used at the expense of cavity transmission bandwidth.   For more control, a gas of known index may be used in combination with a small change in the length of the cavity.

\section{Spectrograph Calibration}
Until now, we have considered only the frequency filtering of comb modes to spectrographically resolvable frequency spacing while maintaining suppression of the neighboring frequency modes.  For this comb to be a useful calibration source, we must also address precision.  Although the constant spacing of the comb lines has been demonstrated, many factors can create an apparent shift of the center of gravity (COG) of the comb lines.  

\begin{figure}[h]
\centering{
\includegraphics*[width=3truein]{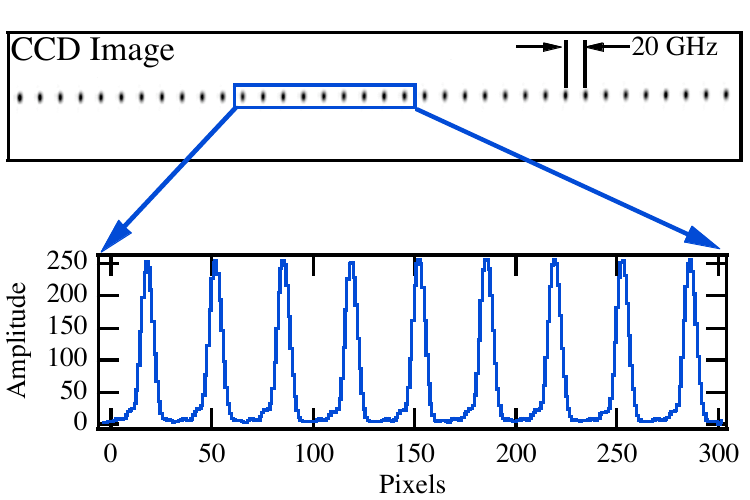}}%
\caption{(a) CCD image of a 1\GHz\ Mode-locked laser filtered to 20\GHz\ by a cavity and then spectrally dispersed with a grating. (b) Intensity plot of subsection.  }
\label{fig_CCD}
\end{figure}

\Fig{fig_CCD} shows an image of 35 comb lines that have been filtered by a 20\GHz\ cavity, dispersed by a high-resolution grating, and imaged onto a charged-coupled device (CCD) camera.  The graph plots a subset of CCD pixels as a function of intensity and shows nine of these comb lines.  For a standard spectrographic calibration, the data would be fit with nine individual Gaussian functions, each with amplitude, width, position, and offset.  
The strength of using a comb to calibrate the spectrograph lies in the ability to leverage the equally spaced frequency components.  Small errors that would typically cause an apparent shift in the center of gravity of one calibration point are now averaged out.   In place of  $N$ Gaussians, each with four free parameters, the entire spectrum may be fit with a sum over $N$ Gaussians, equally spaced along $x$.   Here the sum $\sum_N a{\rm{Exp}}[-c(x-b N-d)^2]+e$ uses only five free parameters for the entire comb:   peak $a$, separation $b$, width $c$, uniform shift $d$, and noise floor $e$.  Because we assume constant spacing, the fit is less sensitive to parameters such as asymmetric suppression of nearest-neighbor modes.   

\begin{figure}[b]
\centering{
\includegraphics*[width=3truein]{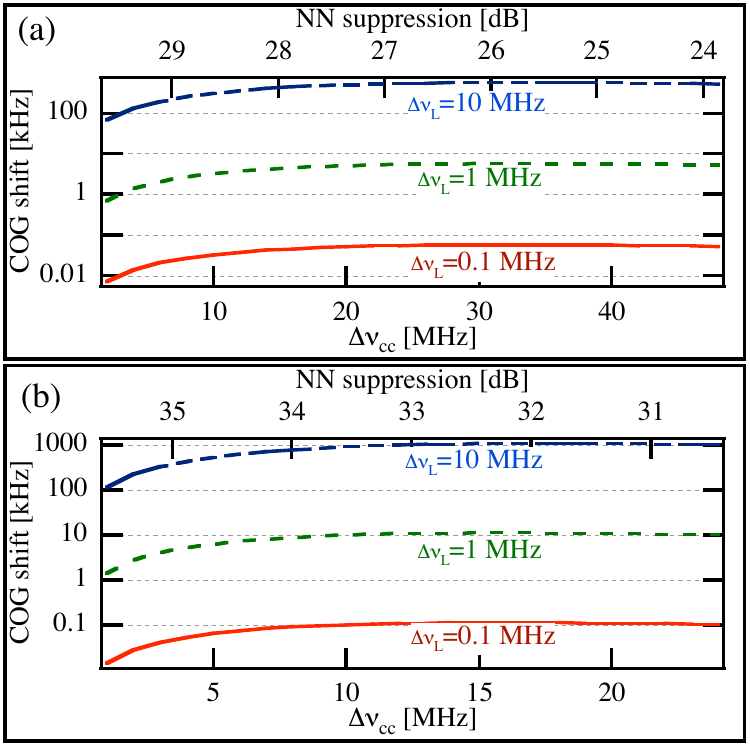}}%
\caption{
Apparent COG shift of a filtered comb line as a function of the offset between the cavity and a comb mode.  Suppression of the nearest neighbor mode is plotted on the upper axis.  Reflectivity of $R=0.99$ is assumed, and the shift is plotted for $\Delta\nu_{\rm L}$ of 0.1\MHz, 1\MHz, and 10\MHz.  Filter numbers of $m=20$  and $m=10$ are  shown in (a) and (b), respectively.  For a $\frep=1\GHz$ comb and $R=0.99$, this corresponds to a cavity linewidth of 64\MHz\ at $m=20$ and 32\MHz\ at $m=10$.  At $\sim$800\nm, a shift of 12.5\kHz\ is equivalent to a radial velocity of 1\,cm/s.}
\label{fig_shift}
\end{figure}

Although calibration in this manner does reduce some systematic errors, the properties of the comb itself must be considered.
In the above simulations, the comb lines were assumed to be infinitely narrow.  Under this premise, the filter cavity acts on the amplitude of a delta function, and no COG shift is possible.  In a physical comb system, the frequency reference, which is used to stabilize the comb, contributes to a linewidth of the comb teeth. When the comb is locked, the frequency modes of the laser may be modeled as a Gaussian with a finite full width at half max, $\nu_{\rm L}$.  If the cavity and comb modes are perfectly overlapped, the comb spectral distribution may slightly change the shape of the comb mode, but does not shift the COG.  As the cavity and comb slip off one another, the cavity asymmetrically reshapes the comb teeth, thereby giving the appearance of a frequency shift.  The larger the ratio of comb tooth width to the cavity width, the greater the shift in COG.  In the limit where the width of the comb tooth exceeds that of the cavity, the cavity function completely reshapes the comb mode, and there is a 1:1 ratio between the COG shift of the comb line and \nucc.
This systematic error would be carried through the spectrograph, creating an apparent shift in calibration.  \Fig{fig_shift} plots the COG shift as a function of the offset between cavity and comb for $m=20$ (a) and $m=10$ (b).  These curves are plotted for a mirror reflectivity of $R=0.99$.  On the upper axis, \Eq{eq_suppress}, is used to denote suppression of the nearest neighbor with respect to the $m^{th}$ comb mode.  The combination of necessary suppression and maximum tolerable shift in COG determine the allowed linewidth of the comb modes.  In a cavity of $R=0.99$ and  $m=10$, which results in a cavity full-width at half-max of $32\MHz$, a comb linewidth of $\Delta\nu_{\rm L}=1\MHz$ has less than 20\kHz\ shift of the COG while suppression is better than 30\dB.  With this level of stabilization, spectroscopic calibrations on the level of a few centimeters per second are possible.
For cavity mirrors with reflectivity greater than 99\,\%,  the shift in COG is greater than that plotted in  \fig{fig_shift}.  

Depending on the level of precision necessary for spectrograph calibration, various frequency reference may be used to stabilize the comb.  A practical and relatively inexpensive option is a low-noise quartz oscillator (typically 5 or 10 MHz) that has its frequency steered on intermediate time scales to a compact microwave atomic standard (such as Rb).  On longer time scales the accuracy of the frequency reference can be guided by signals from the constellation of satellites that form the global positioning system (GPS).  Such frequency references are commercially available and often referred to as GPS-Disciplined oscillators (GPSDO) \cite{Lombardi,Stone}.  The fractional frequency uncertainty available from a GPSDO can be near or below the level of $1\times 10^{-11}$ for averaging times greater that 1 second.  The uncertainty decreases approximately proportional to the averaging time on scales greater than $\sim 1$ day, making feasible high-accuracy measurements even if they are separated by days or years.

It is important not to confuse the frequency uncertainty of the GPSDO with the linewidth it would provide for the frequency comb elements, as we might erroneously assume that an uncertainty of $1\times 10^{-11}$ provides a linewidth of a few kilohertz for the optical comb elements.  In fact, the determination of the exact comb linewidth can involve many factors including, (1) the frequency noise on the free-running femtosecond laser frequency comb, (2) the short-term $(<1\,{\rm s})$ phase-noise of the quartz oscillator in the GPSDO, (3) the residual phase noise on a synthesizer used in converting from the quartz frequency to the repetition rate of the femtosecond laser, and (4) the electronic servo system used in controlling the femtosecond laser comb relative to the GPSDO.  The various contributions of these elements to the observed linewidth of the comb elements can be calculated, keeping in mind that the noise on the quartz crystal and microwave electronics is multiplied by the ratio of the optical and radio/microwave frequencies (\eg, something between $5\times 10^5$ and $5\times 10^7$).  This large multiplication factor can transform the apparently low noise of a 10 MHz oscillator (an the intermediate microwave synthesizer) into an optical linewidth that is much greater than we would naively anticipate.  Nonetheless, with reasonable care and readily available components,  a linewidth of the comb modes that is less than 1 MHz in the visible portion of the spectrum is readily attained.  Improved radio and microwave frequency electronics could further reduce this linewidth.  As just noted, the linewidth scales with frequency, so a difference in linewidth by a factor of $\sim 2$ could be expected in the two extremes of an octave-spanning spectrum.   If a still narrower linewidth for the optical comb modes is required, at the expense of simplicity we could employ a stabilized laser as a short time ($<$ 1 s) reference in place of the quartz oscillator.  In such a case, optical linewidths from ~100 kHz down to $\sim$1 Hz can be achieved~\cite{Bartels04}.

\vspace{.2in}
\section{Conclusion}
Frequency combs are a promising avenue for the calibration of astronomical spectrographs.  Current comb technology requires filtering of comb modes to produce adequate mode separation that can be resolved by standard spectrographs.  The parameters relevant to cavity filtering have been detailed with a demonstrated agreement between experiment and theory, and factors that impact COG line shifts are discussed.
Beyond the compelling benefits of frequency combs as a calibration source for observational astronomy and cosmology, a laser frequency comb
calibration source could positively impact several additional areas that require robust frequency
sources with 10-20 GHz mode spacing. Examples include direct spectroscopy with frequency combs,
remote sensing, optical and microwave waveform synthesis, high-speed coherent communications, and optical
clock development.

The authors gratefully acknowledge Ramin Lalezeri for the data on the mirror coatings used in \fig{fig_Mirror}; Thomas Udem, Ronald Holzwarth, Ted H\"ansch, and Leo Hollberg for sharing their ideas regarding frequency combs for spectrographic calibration; and Andy Weiner for filter cavity design.  Finally, we thank Nathan Newbury and Elizabeth Donley for their thoughtful comments on this manuscript. This work is a contribution of NIST, an agency of  the US government, and is not subject to copyright.


\bibliographystyle{epj} 
\bibliography{CombCavity2} 

\begin{thebibliography}{29}

\bibitem{2007Murphy}
M.~Murphy, T.~Udem, R.~Holzwarth, A.~Sizmann, L.~Pasquini, C.~Araujo-Hauck,
  H.~Dekker, S.~D'Odorico, M.~Fischer, T.W. Hansch et~al., Monthly Notices of
  the Royal Astronomical Society \textbf{380}(2), 839 (2007)

\bibitem{2007Schmidt}
P.O. Schmidt, S.~Kimeswenger, H.U. Kaeufl (2007), \texttt{arXiv:0705.0763
  [astro-ph]}

\bibitem{HallNobel}
J.L. Hall, Reviews of Modern Physics \textbf{78}(4), 1279 (2006)

\bibitem{HanschNobel}
T.W. Hansch, Reviews of Modern Physics \textbf{78}(4), 1297 (2006)

\bibitem{endnote1}
Although this paper is aimed at mature comb technology, it should be noted that
  new advances on femtosecond combs may obviate the need for spectral filtering

\bibitem{Ell01}
R.~{Ell}, U.~{Morgner}, F.X. {K{\"a}rtner}, J.G. {Fujimoto}, E.P. {Ippen},
  V.~{Scheuer}, G.~{Angelow}, T.~{Tschudi}, M.J. {Lederer}, A.~{Boiko} et~al.,
  Optics Letters \textbf{26}, 373 (2001)

\bibitem{2002Bartels}
A.~{Bartels}, H.~{Kurz}, Optics Letters \textbf{27}, 1839 (2002)

\bibitem{Fortier03}
T.M. {Fortier}, D.J. {Jones}, S.T. {Cundiff}, Optics Letters \textbf{28}, 2198
  (2003)

\bibitem{Ranka00}
J.K. {Ranka}, R.S. {Windeler}, A.J. {Stentz}, Optics Letters \textbf{25}, 25
  (2000)

\bibitem{Kumar02}
V.V.R. {Kanth Kumar}, A.K. {George}, W.H. {Reeves}, J.C. {Knight}, P.S.J.
  {Russell}, F.G. {Omenetto}, A.J. {Taylor}, Optics Express \textbf{10}, 1520
  (2002)

\bibitem{Dudley06}
J.M. {Dudley}, G.~{Genty}, S.~{Coen}, Reviews of Modern Physics \textbf{78},
  1135 (2006)

\bibitem{Nicholson03}
J.W. {Nicholson}, M.F. {Yan}, P.~{Wisk}, J.~{Fleming}, F.~{Dimarcello},
  E.~{Monberg}, A.~{Yablon}, C.~{J{\o}rgensen}, T.~{Veng}, Optics Letters
  \textbf{28}, 643 (2003)

\bibitem{Tauser03}
F.~{Tauser}, A.~{Leitenstorfer}, W.~{Zinth}, Optics Express \textbf{11}, 594
  (2003)

\bibitem{Washburn04}
B.R. {Washburn}, S.A. {Diddams}, N.R. {Newbury}, J.W. {Nicholson}, M.F. {Yan},
  C.G. {J{\o}rgensen}, Optics Letters \textbf{29}, 250 (2004)

\bibitem{1999Udem}
T.~Udem, J.~Reichert, R.~Holzwarth, T.W. H\"{a}nsch, Opt. Lett.
  \textbf{24}(13), 881 (1999)

\bibitem{1989Sizer}
T.~Sizer, IEEE J. Quant. Electron. \textbf{25}(1), 97 (1989)

\bibitem{1988DeVoe}
R.G. DeVoe, C.~Fabre, K.~Jungmann, J.~Hoffnagle, R.G. Brewer, Phys. Rev. A
  \textbf{37}(5), 1802 (1988)

\bibitem{2000Jones}
R.J. Jones, J.C. Diels, J.~Jasapara, W.~Rudolph, Optics Communications
  \textbf{175}, 409 (1 March 2000)

\bibitem{Born}
M.~Born, E.~Wolf, \emph{Principles of Optics} (Cambridge University Press,
  1997)

\bibitem{1996Ciddor}
P.E. {Ciddor}, Applied Optics \textbf{35}, 1566 (1996)

\bibitem{1998Abedin}
K.~{Sarwar Abedin}, N.~{Onodera}, M.~{Hyodo}, Applied Physics Letters
  \textbf{73}, 1311 (1998)

\bibitem{2004Yiannopoulos}
K.~{Yiannopoulos}, K.~{Vyrsokinos}, D.~{Tsiokos}, E.~{Kehayas}, N.~{Pleros},
  G.~{Theophilopoulos}, T.~{Houbavlis}, G.~{Guekos}, H.~{Avramopoulos}, IEEE
  Journal of Quantum Electronics \textbf{40}, 157 (2004)

\bibitem{2005McFerran}
J.~McFerran, E.~Ivanov, A.~Bartels, G.~Wilpers, C.~Oates, S.~Diddams,
  L.~Hollberg, Electronics Letters \textbf{41}(11), 650 (26 May 2005)

\bibitem{2008Hollberg}
S.A. Diddams, A.M. Weiner, V.~Mbele, L.~Hollberg, LEOS Summer Topical Meetings,
  2007 Digest of the IEEE pp. 178--179 (23-25 July 2007)

\bibitem{2006Fortier}
T.M. Fortier, A.~Bartels, S.A. Diddams, Opt. Lett. \textbf{31}(7), 1011 (2006)

\bibitem{2007Osterman}
S.~Osterman, S.~Diddams, M.~Beasley, C.~Froning, L.~Hollberg, P.~MacQueen,
  V.~Mbele, A.~Weiner, Proceedings of the SPIE \textbf{6693}, 66931G (2007)

\bibitem{Lombardi}
M.~Lombardi, A.~Novick, V.~Zhang, Frequency Control Symposium and Exposition,
  2005. Proceedings of the 2005 IEEE International pp. 677--684 (29-31 Aug.
  2005)

\bibitem{Stone}
J.~Stone, L.~Lu, P.~Egan, Measure \textbf{2}, 28 (2007)

\bibitem{Bartels04}
A.~Bartels, C.W. Oates, L.~Hollberg, S.A. Diddams, Opt. Lett. \textbf{29}(10),
  1081 (2004)

\end{thebibliography}

\end{document}